\documentclass[aps,pre,amsmath,amssymb,reprint,superscriptaddress]{revtex4-1}
\usepackage{graphicx}
\usepackage{dcolumn}
\usepackage{bm}
\usepackage{color}
\usepackage{enumitem}
\usepackage[normalem]{ulem}
\usepackage{comment}

\begin{document}

\title{The effect of temperature oscillations on energy storage rectification \\ in harmonic systems}

\author{Renai Chen}
\affiliation{Theoretical Division and Center for Nonlinear Studies, Los Alamos National Laboratory, Los Alamos, New Mexico, USA, 87545}

\author{Galen T. Craven}
 \email{galen.craven@gmail.com}
 \affiliation{Theoretical Division, Los Alamos National Laboratory, Los Alamos, New Mexico, USA, 87545}

\begin{abstract}
Rectification, the preferential transport of a current in one direction through a system, has garnered significant attention in molecules because of its importance for controlling thermal and electronic currents at the nanoscale. 
Here, we report the presence of energy storage rectification effects in a molecular chain. This phenomenon is generated by subjecting a harmonic molecular chain to an oscillating temperature gradient and showing that the energy absorption rate of the system depends on the direction of the gradient. We examine how the energy storage rectification ratios in the chain are affected by the oscillating gradient, asymmetry in the chain, and the system parameters. We find that energy storage rectification can be observed in harmonic lattice structures with time-dependent temperatures and that, correspondingly, anharmonicity is not required to generate this rectification mechanism in such systems.
\end{abstract}

\maketitle

\section{Introduction}
Rectification in molecular systems has been well-studied because of its importance in the design of nanoscale technologies 
\cite{Li2004,Wu2007,Kuo2010,Roberts2011,Giazotto2015,Zhao2023, Ratner1974rectifier,Thoss2011,DiezPerez2009,Venkataraman2015,Batista2017}.  
Molecular rectification can be broadly classified into two types: thermal and electronic.
Thermal rectification is a transport phenomenon defined by the preferential flow of heat in one direction through a system \cite{Li2004,He2006,Wu2007,Kuo2010,Roberts2011,Giazotto2015,Zhao2023,RomeroBastida2024}. 
It occurs when a temperature gradient generates a large heat current in one direction through a system, but when the direction of the temperature gradient is reversed, only a small relative amount of heat flows in the opposite direction. 
Thermal rectification is analogous to the electronic rectification processes that occur in electronic diodes \cite{Ratner1974rectifier,Thoss2011,DiezPerez2009,Venkataraman2015,Batista2017}.
Thermal rectifiers/diodes have received significant research interest due to their potential applications in various technologies \cite{Li2004,Wu2007,Kuo2010,Roberts2011,Giazotto2015,Zhao2023,craven2023a,Segal2005jcp,Segal2008prl,Segal2009,Leitner2013,craven18b,Simon2021,Krivtsov2023}.
Electronic rectification is fundamental in many electrical circuit designs
where it is used to control the directionality of electrical currents and to switch a system between different states for logical operations and information processing. 

Molecular thermal rectification---the preferential flow of heat in one direction through a nanoscale molecular system---is a complex phenomenon that can involve multiple heat transport processes including vibrational, electronic, and radiative mechanisms, 
and the interplay between them \cite{craven2023a,Segal2005jcp,Wu2007,Wu2009,Segal2008prl,Segal2009,Leitner2013,craven18b,Simon2021}.
Molecular thermal rectification falls under the broad umbrella of nanoscale energy transport, a research area with broad importance due to its relevance in the design of new nanotechnologies \cite{Li2012,Segal2016,Sabhapandit2012,Lebowitz1959,Lebowitz1967,Lebowitz1971,Lebowitz2008,Nitzan2003thermal,Segal2005prl,Lebowitz2012,Dhar2015,Velizhanin2015,Esposito2016,craven16c,matyushov16c,craven17a,craven17b,craven17e,craven18b,craven20a,craven21a,chen2020local,chen2023jcp,Ochoa2022}.
Theoretical descriptions of nanoscale and molecular energy transport have been actively developed for decades \cite{Lebowitz1959,Lebowitz1967,Lebowitz1971,Cahill2003,Segal2005prl,Segal2016,Nitzan2007,Sato2012,Maldovan2013,Leitner2008,Leitner2013,Seifert2015periodictemp,Li2012,Dubi2011,Lim2013,He2021,Volz2022, HernandezJPCL2023, Krivtsov2023}. 
These studies have advanced our fundamental understanding of nanoscale energy transport and how it can be utilized to design new nanotechnologies and devices \cite{Li2006,Ben-Abdallah2014,Joulain2016,Wang2017thermaldiode,craven17a,craven18b}.

It has recently been observed that subjecting a system to a time-periodic temperature gradient can alter the energy transport properties of the system in comparison to a static temperature gradient 
and give rise to novel and emergent transport phenomena \cite{urban2022thermal,craven2023b,craven2024a}.
Time-dependent temperatures can be used 
to affect system properties across various length scales \cite{goss1998quantitative,platkov2014periodic,singleatom2016, Hern2007} including to induce thermal rectification in 
macroscopic solid-state systems \cite{Shimokusu2022}.
Temperature modulations play an important role in the function of a multitude of systems including
energy storage technologies \cite{Gur2012thermalbattery,Wang2022battery},
energy harvesting materials \cite{Bowen2014pyro,Yamamoto2021,Lheritier2022pyro},
heat transport devices \cite{bartussek1994periodically,Hanggi2009wireperiodictemps,Zhang2008ratchet},
and thermal logic devices with memory  \cite{Ben-Abdallah2017thermalmemristor,Ordonez-Miranda2019thermalmemristor}.
Therefore, developing theoretical tools to describe these processes and using those tools to discover new energy transport mechanisms is important in a diverse range of research fields.

In this article, we report a rectification mechanism 
that differs from thermal or electronic rectification.
This phenomenon is termed \textit{energy storage rectification}---an effect in which the amount of energy stored by a system depends on the direction of an applied thermal gradient. 
We specifically examine energy storage rectification and how it can induced and then controlled using a temperature gradient that is oscillating in time.
We present results that illustrate how an oscillating temperature gradient affects energy storage rectification ratios in a harmonic molecular chain. We apply 
the formalism developed previously by us in Ref.~\citenum{craven2023b} and Ref.~\citenum{craven2024a} to calculate the energy fluxes and the corresponding energy storage properties in the model.
This formalism uses a nonequilibrium Green's function approach adapted to treat the non-stationary distributions that arise when the bath temperatures are time-periodic. Here, this formalism is applied to understand energy storage in harmonic chains.
We calculate the energy flux in/out of the chain under forward and reverse time-periodic thermal bias conditions and then compare their ratio. 
We find that energy storage rectification effects can be observed in harmonic chains and that, correspondingly, 
anharmonicity is not required to observe these effects \cite{Terraneo2002,Segal2009SCR,Ming2016,Simon2021}. 

\section{Model}

The model we consider is a one-dimensional harmonic chain of $N$ particles connecting two heat baths with temperatures that are oscillating in time. The two heat baths are labelled L for ``left'' and R for ``right'', respectively.
The equations of motion for this system are:
\begin{equation}
\begin{aligned}
\label{eq:EoM1}
m_1 \ddot{x}_1 &= -k x_1 + k x_2 - m_1 \gamma_\text{L} \dot{x}_1 + \xi_\text{L}(t), \\
 \ldots \\
m_2 \ddot{x}_j  &= -2kx_j+kx_{j-1}+kx_{j+1}, \\
 \ldots \\
 m_N \ddot{x}_N &= -k x_N + k x_{N-1} - m_N \gamma_\text{R} \dot{x}_N + \xi_\text{R}(t),
\end{aligned}
\end{equation}
where $m_j$ is the mass of the $j$th particle, $x_j$ is the displacement of particle $j$ from its equilibrium position,
$k$ is the harmonic force constant between particles,
$\gamma_\text{L}$ and $\gamma_\text{R}$ are coupling constants between the system and the respective baths, and 
$\xi_\text{L}(t)$ and $\xi_\text{R}(t)$ are stochastic forces that obey the correlations
\begin{equation}
\begin{aligned}
\label{eq:FD_theorem_t}
 \big\langle \xi_\text{L}(t) \xi_\text{L}(t')\big\rangle &= 2 \gamma_\text{L}m k_\text{B} T_\text{L}(t)\delta(t-t'), \\
 \big\langle \xi_\text{R}(t) \xi_\text{R}(t')\big\rangle &= 2 \gamma_\text{R}m k_\text{B} T_\text{R}(t)\delta(t-t'), \\
	\big\langle \xi_\text{L}(t) \xi_\text{R}(t')\big\rangle &= 0, \\
 \big\langle \xi_\text{L}(t)\big\rangle &=0,\\
 \big\langle \xi_\text{R}(t)\big\rangle &=0,
\end{aligned}
\end{equation}
where $k_\text{B}$ is the Boltzmann constant and $T_\text{L}(t)$ and $T_\text{R}(t)$ are the time-dependent and oscillatory temperatures of the left and right bath, respectively. 
Details on the specific functional forms we use for the time-dependent temperatures are given below.
It is important to note that the harmonic system in Eq.~(\ref{eq:EoM1}) with correlations in Eq.~(\ref{eq:FD_theorem_t}) does not exhibit thermal rectification effects, even in the presence of temperature oscillations \cite{Terraneo2002,Segal2009SCR,Ming2016,Simon2021,Segal2021}. Our aim in this article is to examine if energy storage rectification effects can be generated in a harmonic system by applying an oscillatory temperature gradient. 

\begin{figure*}
    \centering
	\includegraphics[width=\textwidth]{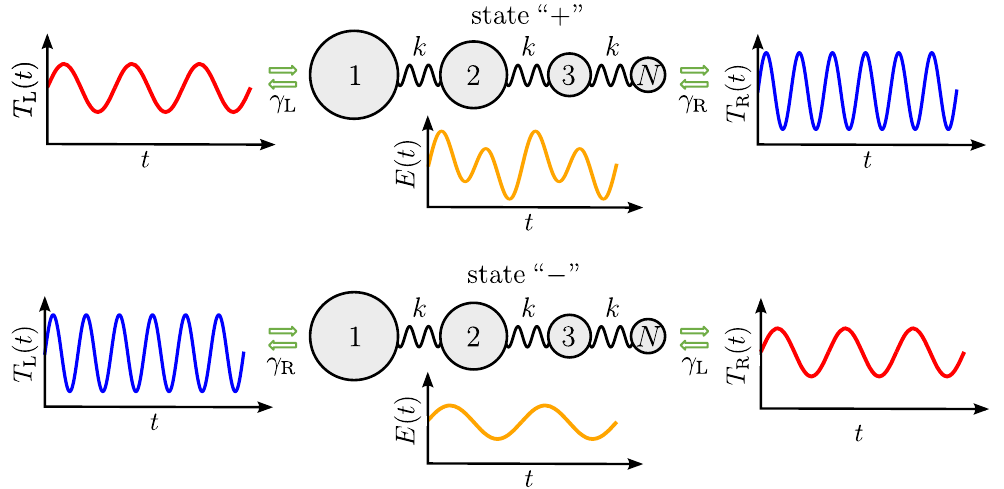}
	\caption{Schematic diagram of a molecular chain connecting two heat baths with oscillating temperatures
    in the baseline $``+"$ (top) and reverse $``-"$ (bottom) thermal bias states.
    The temperatures of the left bath and right bath, $T_\text{L}(t)$ and $T_\text{R}(t)$, are oscillating in time as illustrated by the graphs on left and right of the figure.
    In the reverse state, the temperatures of the two baths are swapped: $T_\text{L}(t)\rightarrow T_\text{R}(t)$ and $T_\text{R}(t)\rightarrow T_\text{L}(t)$. The corresponding system-bath couplings $\gamma_\text{L}$ and $\gamma_\text{R}$ are also swapped.
    The system energy flux $\partial_t E(t)$ is different in each thermal bias state as illustrated by the graphs in the middle of the figure.
    }
	\label{fig:schematic}
\end{figure*}

To examine energy storage rectification, we consider two states for the time-dependent thermal gradient between heat baths: a baseline state denoted by ``$+$'' and a state with the temperature gradient reversed denoted by ``$-$''. 
In state ``+'',  the bath temperatures take the specific forms:
\begin{equation}
\begin{aligned}
\label{eq:t-dep_T_plus}
 T^+_\text{L}(t)   &= T^{(0)}_\text{L} + \Delta T_\text{L} \sin(\omega_\text{L} t),\\[1ex]
 T^+_\text{R}(t)   &= T^{(0)}_\text{R} + \Delta T_\text{R} \sin(\omega_\text{R} t),
\end{aligned}
\end{equation}
where $T^{(0)}_\text{L}$ and $T^{(0)}_\text{R}$ are the temperatures of the two baths in the limit of vanishing of temperature oscillations, $\Delta T_\text{L} $ and $\Delta T_\text{R}$ are the amplitudes of the oscillations, and $\omega_\text{L}$ and $\omega_\text{R}$ are oscillation frequencies. 
The system-bath couplings are $\gamma_\text{L}$ and $\gamma_\text{R}$
and we define $\gamma = \gamma_\text{L}+\gamma_\text{R}$.
In this article, we will always consider cases in which $\omega_\text{L}$ and $\omega_\text{R}$ are commensurate so that the system is periodic with total period $\mathcal{T}$.
In the baseline state, the temperature difference between the two baths is 
\begin{equation}
\Delta T^+ (t) = T^+_\text{R}(t) - T^+_\text{L}(t).
\end{equation}
and the noise correlations are
\begin{equation}
\begin{aligned}
\label{eq:FD_theorem_t_pos}
 \big\langle \xi^+_\text{L}(t) \xi^+_\text{L}(t')\big\rangle &= 2 \gamma_\text{L} m_1 k_\text{B} T^+_\text{L}(t)\delta(t-t'), \\
 \big\langle \xi^+_\text{R}(t) \xi^+_\text{R}(t')\big\rangle &= 2 \gamma_\text{R} m_N k_\text{B} T^+_\text{R}(t)\delta(t-t').
\end{aligned}
\end{equation}
In state ``$-$'', the temperature gradient is reversed with
\begin{equation}
\begin{aligned}
\label{eq:t-dep_T_minus}
 T^-_\text{L}(t)   &= T^{(0)}_\text{R} + \Delta T_\text{R} \sin(\omega_\text{R} t),\\[1ex]
 T^-_\text{R}(t)   &= T^{(0)}_\text{L} + \Delta T_\text{L} \sin(\omega_\text{L} t),
\end{aligned}
\end{equation}
and the temperature difference between the baths is
\begin{equation}
\Delta T^- (t) =  -\Delta T^+ (t).
\end{equation}
The noise correlations in the reverse thermal bias state are
\begin{equation}
\begin{aligned}
\label{eq:FD_theorem_t_minus}
 \big\langle \xi^-_\text{L}(t) \xi^-_\text{L}(t')\big\rangle &= 2 \gamma_\text{R} m_1 k_\text{B} T^-_\text{L}(t)\delta(t-t'), \\
 \big\langle \xi^-_\text{R}(t) \xi^-_\text{R}(t')\big\rangle &= 2 \gamma_\text{L} m_N k_\text{B} T^-_\text{R}(t)\delta(t-t').
\end{aligned}
\end{equation}
Note that the system-bath couplings $\gamma_\text{L}$ and $\gamma_\text{R}$ are also reversed in this state. 
Figure~\ref{fig:schematic} is a schematic diagram of the model in the two thermal bias states. 

The oscillating temperature gradient gives rise to time-dependent energy fluxes in the applied model.
Those energy fluxes can be separated into three terms:
(1) $J_\text{sys}$ is the energy flux in/out of the system, i.e., the molecular chain, (2) $J_\text{L}$ is the energy flux associated with the left bath 
and (3) $J_\text{R}$ is the energy flux associated with the right bath. 
In the limit of constant bath temperatures, the system will reach a steady state in which the system energy flux vanishes, $J_\text{sys}=0$. But, in the systems examined here, the system energy flux does not vanish because of the temperature oscillations \cite{craven2023a,craven2024a}.
The system energy flux in the respective thermal bias state is 
\begin{equation}
\label{eq:general_heat}
J^\pm_\text{sys} (t) = \partial_t \big\langle E(t)\big\rangle^\pm, 
\end{equation}
where
\begin{equation}
E(t) = \sum_{i = 1}^N\frac{1}{2} m_i \dot{x}^2_i(t) + \sum_{i = 1}^{N-1} \frac{1}{2}k\left[x_i(t) - x_{i+1}(t)\right]^2,
\end{equation}
is the total energy of the chain,
$\langle  \rangle$ represents an ensemble average,
and the superscripts $\pm$ denote that the expectation value is evaluated in the corresponding bias state.
The energy fluxes for the left and right baths in each bias state are
\begin{align}
\label{eq:heatcurrentbathL}
J^+_\text{L} (t) &=  m \gamma_\text{L} \big\langle \dot{x}_1^2(t)\big\rangle^+ - \big\langle\xi_\text{L}(t) \dot{x}_1(t)\big\rangle^+,\\[1ex]
\label{eq:heatcurrentbathR}
J^+_\text{R} (t) &=  m \gamma_\text{R} \big\langle \dot{x}_N^2(t)\big\rangle^+ - \big\langle \xi_\text{R}(t) \dot{x}_N(t)\big\rangle^+,\\[1ex]
\label{eq:heatcurrentbathLminus}
J^-_\text{L} (t) &=  m \gamma_\text{R} \big\langle \dot{x}_1^2(t)\big\rangle^- - \big\langle\xi_\text{L}(t) \dot{x}_1(t)\big\rangle^-,\\[1ex]
\label{eq:heatcurrentbathRminus}
J^-_\text{R} (t) &=  m \gamma_\text{L} \big\langle \dot{x}_N^2(t)\big\rangle^- - \big\langle \xi_\text{R}(t) \dot{x}_N(t)\big\rangle^-.
\end{align}
The functional forms for these expressions can be derived using a stochastic energetics formalism \cite{Sabhapandit2012,Sekimoto1998}.
Because the bath temperatures are periodic in time, 
the model will not reach steady state defined by $\partial_t \big\langle E(t)\big\rangle = 0$ and $J_\text{L} (t) = -J_\text{R} (t)$. 
Instead, it approaches a time-dependent nonequilibrium state 
with an average energy that is oscillating in time and, therefore, a system energy flux $J_\text{sys} (t)$ that is not equal to zero for all $t$. The oscillating system energy implies that the system is storing and releasing energy as the temperature gradient is oscillating. Here, we address the question: Does the direction of the time-periodic temperature gradient affect energy storage and release?

To quantify the extent of energy storage rectification, we calculate the energy absorbed by the system over a period of oscillation $\mathcal{T}$ in each of the two thermal bias states ``$+$'' and ``$-$'' and compare them. 
The energy stored by the system over one period of oscillation is \cite{craven2023a,craven18a1,craven18a2,BaratoPRL2018}
\begin{equation}
\label{eq:heatplussys}
\mathcal{Q}^\pm_\text{storage} =  \int_0^{\mathcal{T}} J^\pm_\text{sys} (t') \Theta[J^\pm_\text{sys} (t') ]  dt'
\end{equation}
where $\Theta$ is the Heaviside function. 
We term this quantity the energy storage capacity.
It is important to note that over a period of oscillation, the total system energy storage is offset by the total energy release, meaning that $\int_0^{\mathcal{T}} J^\pm_\text{sys} (t') dt' = 0$.
The energy storage rectification ratios are defined using 
\begin{equation}
\label{eq:ratio}
R_\text{storage} = \frac{\mathcal{Q}^-_\text{storage} }{\mathcal{Q}^+_\text{storage} } = \frac{\int_0^{\mathcal{T}} J^-_\text{sys} (t')  \Theta[J^-_\text{sys} (t') ] dt'}{\int_0^{\mathcal{T}} J^+_\text{sys} (t') \Theta[J^+_\text{sys} (t') ] dt'}.
\end{equation}
When $R_\text{storage} \neq 1$, energy storage rectification is observed.

To generate asymmetry in the system (a property that is necessary for rectification) we vary the masses along the chain. 
Specifically, we use two different models, Model I and Model II, for the masses in Eq.~(\ref{eq:EoM1}).  
In Model I, we generate a mass gradient by taking the mass of the $i$th particle in the chain to be $m_i = m_\text{max} - (i-1)(m_\text{max} - m_\text{min})/(N-1)$ where $m_\text{max}$ is the mass of particle $1$ and $m_\text{min}$ is the mass of particle $N$
\cite{Li2007massgradient} with $m_\text{max}>m_\text{min}$.
In Model II, each particle $i$ in the chain has the same mass $m_i = m$ except for a single massive particle at position $n$ that has mass $m_n \gg m_i \, \forall \, i \neq n$.
Therefore, Model I describes a chain with a mass gradient and Model II describes a chain with single point mass disorder. Both models lead to mass asymmetry.

\section{Results}

We have previously developed a formalism in Ref.~\citenum{craven2024a}  to calculate the energy fluxes in a harmonic chain using 
a nonequilibrium Green's function technique adapted to treat the non-stationary distributions that arise when the bath temperatures are time-periodic. 
Here, for brevity, we will apply those results without details or mathematical exposition, but for a complete description of the applied theoretical approach we refer the reader to our previous work \cite{craven2024a}.



\begin{figure}[t]
\centering
\includegraphics[width = \columnwidth,clip]{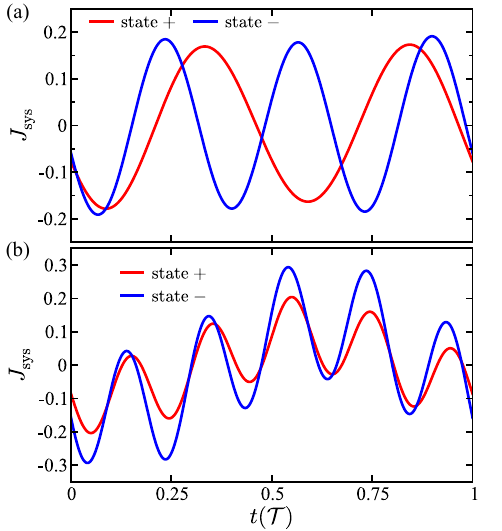}
\caption{\label{fig:sys_energy_time}
Time-dependent system energy fluxes $J_\text{sys}$ as a function 
of time for Model I with different sets of parameters. 
Results for the baseline ($+$) and and reverse ($-$) thermal bias states are shown in red and blue, respectively.
Time is shown in units of the total oscillation period $\mathcal{T}$.
In (a), the parameters are $N = 5$, $m_\text{max} = 10$, $m_\text{min} = 1$, $\gamma_\text{L} = \gamma_\text{R} = 2$, $k = 25000$, $T^{(0)}_\text{L} = 1.5$, $T^{(0)}_\text{R} = 1$, $\Delta T_\text{L} = \Delta T_\text{R} = 0.1$, $\omega_\text{L} =2$, $\omega_\text{R} = 3$.
In (b), the parameters are  $N = 2$, $m_\text{max} = 10$, $m_\text{min} = 1$, $\gamma_\text{L} = \gamma_\text{R} = 2$, $k = 250$, $T^{(0)}_\text{L} = 1.5$, $T^{(0)}_\text{R} = 1$, $\Delta T_\text{L} = \Delta T_\text{R} = 0.1$, $\omega_\text{L} =2$, $\omega_\text{R} = 10$.
All parameters throughout this paper are given in reduced units with 
characteristic dimensions: $\widetilde{\sigma} = 1\,\text{\AA}$,  $\widetilde{\tau} = 1/\widetilde{\gamma} = 1\,\text{ps}$,
$\widetilde{m} = 10\,m_u$ and $\widetilde{T} = 300\,\text{K}$.
}
\end{figure}

Figure~\ref{fig:sys_energy_time} shows the system energy flux $J_\text{sys}(t)$ 
in each of the two thermal bias states (baseline and reverse) as function of time for two sets of system parameters. 
The values of $J^+_\text{sys}(t)$ and $J^-_\text{sys}(t)$ are represented respectively by the red and blue curves.
The results in Fig.~\ref{fig:sys_energy_time} (a)
are for a chain with $N = 2$ particles where one of the particles has more mass than the other (Model I).
This mass asymmetry in the chain generates a mass gradient.
The temperature oscillation frequencies are $\omega_\text{L} =2$ and $\omega_\text{R} = 3$.
It can be seen that $J^+_\text{sys}(t)$ has a different functional form than $J^-_\text{sys}(t)$.
In this case, the oscillation phase of the system energy can be significantly different in the two bias states. 
This implies that the energy storage in each of the two states $\mathcal{Q}^\pm_\text{storage}$
will be different, i.e., that energy storage rectification effects could be observed.
Results for a chain of $N = 5$ particles are shown in Fig.~\ref{fig:sys_energy_time} (b).
In this case, only the left bath temperature is oscillating,  $\omega_\text{L} =5$, while the other bath temperature is constant.
Again, $J^+_\text{sys}(t)$ is different from  $J^-_\text{sys}(t)$.
The primary observation in Fig.~\ref{fig:sys_energy_time} is that the system energy flux depends on the thermal bias state, i.e., the direction of the thermal gradient, a phenomenon that gives rise to energy storage rectification effects.
It is important to note that we do not observe any thermal rectification effects in any of the models examined in this manuscript.

As illustrated in Fig.~\ref{fig:sys_energy_time}, asymmetric chains can show differences in peak magnitudes and functional forms for the system energy fluxes in the different bias states.
In a symmetrical chain, however, the system energy flux will be the same in both bias states, $J^+_\text{sys}(t)=J^-_\text{sys}(t)$, and there will be no energy storage rectification. 
We have confirmed that the numerical codes used to generate the results presented here give this result for the case of a chain with no mass asymmetry.
Also, note that rectification will only be observed away from the quasistatic limit
which is the limit in which the baths oscillate slowly with respect to the system-bath couplings  ($\omega_\text{L}/\gamma, \omega_\text{R}/\gamma \rightarrow 0,0)$.
In the quasistatic limit, $J^\pm_\text{sys} = 0$ and so $R_\text{storage} \to 1$, so there is no rectification in this limit.

\begin{figure}[t]
\includegraphics[width = 8.5cm,clip]{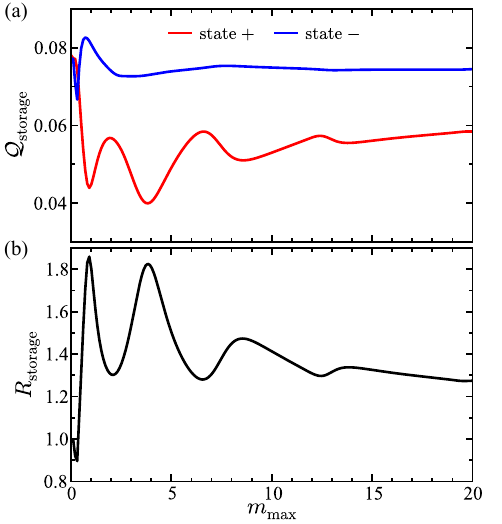}
\caption{\label{fig:rectification1}
(a) Energy storage capacity of Model I calculated as a function mass gradient quantified by varying $m_\text{max}$ while keeping $m_\text{min} = 0.1$ constant. The chain length is $N = 5$.
The energy storage is calculated using Eq.~\ref{eq:heatplussys} and shown in units of $k_\text{B} \widetilde{T}$.
(b) The corresponding energy storage rectification ratio as a function of $m_\text{max}$.
Parameters are $\gamma_\text{L} = \gamma_\text{R} = 2$, $k= 250$, $T^{(0)}_\text{L} = 1.5$, $T^{(0)}_\text{R} = 1.0$,
$\Delta T_\text{L} = 0.1$, $\Delta T_\text{R} = 0$, $\omega_\text{L} =5$, and $\omega_\text{R} = 0$.
All values are given in reduced units as specified in the caption of Fig.~\ref{fig:sys_energy_time}.
}
\end{figure}

Figure~\ref{fig:rectification1} shows the energy storage results for Model I (a chain with a mass gradient) as a function of increasing mass gradient magnitude for a chain with $N =5$ particles. 
Varying the maximum mass $m_\text{max}$ while holding $m_\text{min} = 0.1$ constant gives rise to  nonlinear behaviors in the energy storage capacities, as shown in Fig.~\ref{fig:rectification1}(a).
The energy storage capacities in both the baseline and reverse bias states are nonmonotonic and nonlinear with respect to variation of the mass gradient magnitude. 
This gives rise to  behaviors in which the rectification ratio oscillates as the mass gradient is varied, as shown in Fig.~\ref{fig:rectification1} (b). This result implies there are frequency-dependent resonances in the 
phononic density of states that are more strongly influenced in one of the bias states than the other.
We observe maximum energy storage rectification ratios up to $\approx 1.8$ in this system. 
This illustrates that by reversing the thermal bias state in a system with oscillating temperatures, the energy storage capacity can be significantly different in one state compared to the other.



\begin{figure}[t]
\includegraphics[width = 8.5cm,clip]{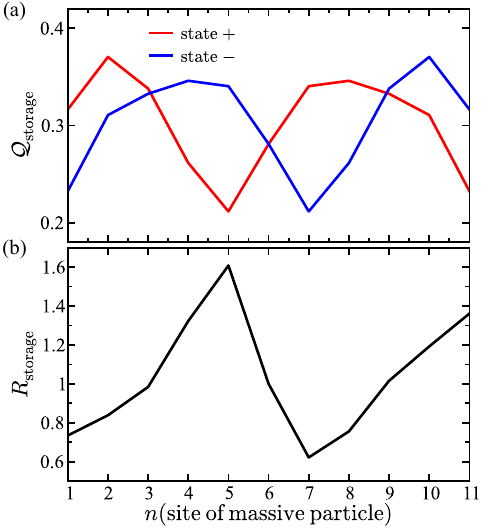}
\caption{\label{fig:rectificationN}
(a) Energy storage capacity of Model II as a function 
of single-point mass disorder. 
Each particle in the chain has a mass of $m = 1$ except for a single massive particle at location $n$ (shown on the $x$-axis) with $m_n=10$.
The values of $\mathcal{Q}^+_\text{storage}$ are shown in red and $\mathcal{Q}^-_\text{storage}$ in blue.
(b) Energy storage rectification ratio for the same system as a function of the location of the massive particle in the chain.
Parameters in both panels are $\gamma_\text{L} = \gamma_\text{R} = 2$, $k = 250$, $T^{(0)}_\text{L} = 1$, $T^{(0)}_\text{R} = 1.5$,
$\Delta T_\text{L} = \Delta T_\text{R} = 0.1$, $\omega_\text{L} =3$, and $\omega_\text{R} = 2$.
}
\end{figure}

Shown in Fig.~\ref{fig:rectificationN}(a) are the values of $\mathcal{Q}^\pm_\text{storage}$ in
Model II, a chain with a single point mass disorder. 
In this model, there is one massive particle in a chain of particles with otherwise uniform masses. 
The $x$-axis shows
the position of the massive particle. It can be observed that the energy storage values do not change smoothly with variation of the site of the mass disorder, instead oscillating (zig-zag) patterns are observed.
This means a small alteration of mass disorder can change significantly change the energy storage capacity in the two thermal bias states. 
The energy storage capacity of both bias states is symmetric for $n=6$. This is expected because this is the case 
when the massive particle is located in  middle of the chain. 
The system is symmetric in this case and so the energy storage capacity is the same both states.
The corresponding rectification ratios are shown in Fig.~\ref{fig:rectificationN} (b).
Note that the examined system is completely harmonic and that no anharmonicities exist in the particle-particle interactions.
The rectification ratios vary between approximately 0.6 and 1.6 depending on the location of the mass disorder in the chain. 
When the massive particle is located on the exact middle of the chain $n=6$,
there is no rectification ($R_\text{storage} = 1$) as expected for a symmetric system.
A related observation is that it is not necessarily true that the rectification is smaller when the mass disorder is closer to the middle of the chain. In fact, the highest rectification value occurs when the massive particle is located at the two positions ($n=5$ and $n=7$) adjacent to the middle symmetric site. This means that the rectification effects cannot be generally correlated with the degree of deviation of the center-of-mass of the chain from the symmetrical position.

\begin{figure}[t]
\includegraphics[width = 8.5cm,clip]{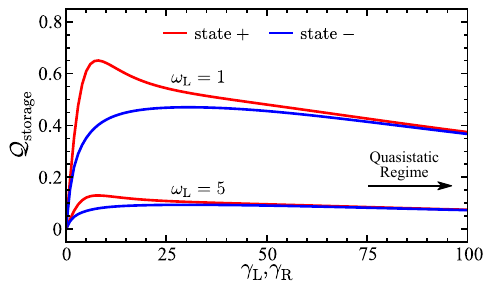}
\caption{\label{fig:RectificationGamma}
Energy storage capacity as a function of the system-bath couplings $\gamma_\text{L}$ and $\gamma_\text{R}$ with 
$\gamma_\text{L}=\gamma_\text{R}$ for Model II.
The values of $\mathcal{Q}^+_\text{storage}$ are shown in red and $\mathcal{Q}^-_\text{storage}$ in blue.
The top set of curves are for $\omega_\text{L} = 1$ and the bottom set of curves are for $\omega_\text{L} = 5$.
The mass of each particle in the chain is $m = 1$ except $m_2=100$.
Other parameters are $N=10$, $k = 250$, $T^{(0)}_\text{L} = 1$, $T^{(0)}_\text{R} = 1.5$,
$\Delta T_\text{L} = 0.1$, $\Delta T_\text{R} = 0$, and $\omega_\text{R} = 0$ .
All units are given in reduced units as specified in the caption of Figure~\ref{fig:sys_energy_time}.
}
\end{figure}

Figure~\ref{fig:RectificationGamma} illustrates how the energy storage capacity in each bias state changes as the 
system-bath coupling strengths are varied for a model with single point mass disorder (Model II). 
Here, we take the couplings for each bath to be equal, $\gamma_\text{L} = \gamma_\text{R}$, 
so the $x$-axis represents the value of both couplings simultaneously. 
The values of $\mathcal{Q}^+_\text{storage}$ are shown in red and $\mathcal{Q}^-_\text{storage}$ in blue,
and results are shown for two different temperature oscillation frequencies as labeled in the plot.
The energy storage capacities can be significantly altered by changing the system-bath coupling strengths.

There are three important physical regimes that can be observed.
As $\gamma_\text{L}$ and $\gamma_\text{R}$ go to zero, 
the coupling between the chain and the thermal baths is too small to facilitate the amount of energy transfer from the baths to the system 
needed to observe significant energy storage. In this regime, $\mathcal{Q}^\pm_\text{storage} \to 0$.
In another regime, as $\gamma_\text{L}$ and $\gamma_\text{R}$ are increased from zero, the energy storage increases, goes through a maximum
and then begins to decrease. This turnover behavior is analogous to Kramer's turnover behavior
which is observed in chemical reaction rates and thermal conductance properties \cite{hern08g,Nitzan2010, Velizhanin2015}.
As $\gamma_\text{L}$ and $\gamma_\text{R}$ become large, $\mathcal{Q}^\pm_\text{storage} \to 0$. This is the quasistatic limit in which $J^\pm_\text{sys} = 0$ due to slow temperature oscillations relative to the system-bath coupling frequencies. 
In the quasistaic regime, the system dissipates energy into the baths on a time scale much faster than the temperature oscillation frequency, and therefore the system energy flux vanishes. 
The rectification ratios for this system exhibit a similar turnover behavior with
$R_\text{storage} \to 1$ as the system-bath couplings go to 0 and $R_\text{storage} \to 1$ in the opposite limit of strong system-bath coupling as well. 
In between these two limiting regimes, the energy storage rectification effects will be maximized.

\begin{figure}[t]
\includegraphics[width = 8.5cm,clip]{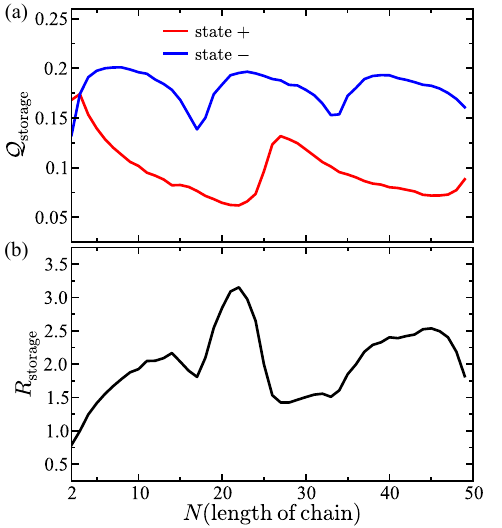}
\caption{\label{fig:RectificationN}
(a) Energy storage capacity of Model II as a function of chain length $N$.
Each particle in the chain has a mass of $m = 1$ except for the first particle which has mass $m_1=100$.
The values of $\mathcal{Q}^+_\text{storage}$ are shown in red and $\mathcal{Q}^-_\text{storage}$ in blue.
(b) Energy storage rectification ratio for the same system as a function of chain length.
Parameters in both panels are $\gamma_\text{L} = \gamma_\text{R} = 2$, $k = 2500$, $T^{(0)}_\text{L} = 1$, $T^{(0)}_\text{R} = 1.5$,
$\Delta T_\text{L} = \Delta T_\text{R} = 0.1$, $\omega_\text{L} =3$, and $\omega_\text{R} = 2$.}
\end{figure}

The rectification effects can also be altered by varying the length of the chain. 
Figure~\ref{fig:RectificationN} illustrates this behavior in Model II. 
The energy storage capacities shown in Fig.~\ref{fig:RectificationN}(a) show that
different chains lengths have different storage capacities and that the effect of varying the chain length is different in each bias state.
Figure~\ref{fig:RectificationN}(b) shows the corresponding rectification ratios. 
The primary observation is that rectification varies nonmonotonically and nonlinearly as the chain length is varied. 
This behavior can be attributed to the time-dependent populations of vibrational modes in the system due to
the temperature oscillations. 
In essence, the oscillating temperatures 
induce a time-dependent power spectrum in which the system's vibrational modes populate and depopulate over time by factors that are
nonlinearly proportional chain length.
Varying the chain length can also significantly affect the magnitude of the rectification ratio with values up to $\approx$3.0 observed for this set of system parameters.

\noindent Overall, several observations are of note:
\begin{itemize}
\item Energy storage rectification effects can be generated due to temperature gradient oscillations.
\item We do not observe thermal rectification.
\item There must be a system energy flux to observe energy storage rectification.  This only occurs away from the quasistatic limit, i.e.,  the limit in which $J^\pm_\text{sys} = 0$ due to slow temperature oscillations relative to the system-bath coupling frequencies. Rectification will only be observed away from  this limit.
\item A turnover behavior is observed in the energy storage rectification with respect to variation of the system-bath couplings. This effect is similar to Kramer's turnover.
\item We have illustrated the rectification effects over a limited parameter range. A comprehensive study of how each parameter in the model affects rectification is an important next step. 
\end{itemize}

\section{\label{conc} Conclusions}

We have examined energy storage rectification in a harmonic molecular chain in the presence of a 
temperature gradient that is oscillating in time.
The presented results illustrate how an oscillating temperature gradient modifies energy transport through a molecular lattice leading to rectification effects.
These effects arise due to differences between the system-bath relaxation rates and other frequencies in the system, for example the oscillation frequency of the temperature gradient, that generate non-stationary distributions in the molecular chain.
We do not observe any net thermal rectification effects in the applied harmonic model, 
however future work on anharmonic systems is an important next step in this direction.
We have demonstrated that harmonic systems can exhibit energy storage rectification effects. 
Overall, the time-periodic modulation of a temperature gradient can affect the energy absorbing properties of nanoscale and molecular systems and give rise to emergent phenomena.
Our results open a new design strategy for molecular devices, capacitors, and batteries with energy storage properties and power cycles that are controlled using time-dependent temperature oscillations.

\section{Acknowledgments}
We acknowledge support from the Los Alamos National Laboratory (LANL) 
Directed Research and Development funds (LDRD).
This research was performed in part at the Center for Nonlinear Studies (CNLS) at LANL. 
The computing resources used to perform this
research were provided by the LANL Institutional Computing Program.

\bibliographystyle{apsrev}

\end{document}